\documentclass{pasa}%

\title[OSC: ESO 65 03]{Complex stellar system ESO65SC03: Open cluster or remnant?}
\author[Joshi et al.]{Gireesh C. Joshi{$^1$}\thanks{gchandra.2012@rediffmail.com}, Y. C. Joshi{$^2$}, S. Joshi{$^2$}, S. Chowdhury{$^2$} and R. K. Tyagi{$^3$}\\
\affil{$^1$Center of Advanced Study, Department of Physics, D.S.B. Campus, Kumaon University, Nainital-263002}
\affil{$^2$Aryabhatta Research Institute of Observational Sciences (ARIES), Manora peak, Nainital, India}
\affil{$^3$Department of Physics, Hemwati Nandan Bahuguna Govt. P.G. College,
Khatima-262308, Uttarakhand}}%
\jid{PASA}
\doi{10.1017/pas.\the\year.xxx}
\jyear{\the\year}

\begin{document}%
\begin{abstract}
We present a complete spatial and dynamical study of the poorly populated stellar system ESO65SC03. The radial distribution of the system gives a core and cluster radii of $1.10{\pm}0.63$ arcmin and 5.36${\pm}$ $0.24$ arcmin, respectively. The surface number density profile (SNDP) does not show any clear enhancement of the surface stellar number density between the stars of the system and the field regions. We derive the optimum isochrone solution for a particular grid size in the colour-magnitude diagram (CMD) using the statistical cleaning procedure. Using the statistically cleaned CMDs, we find the distance modulus, $(m-M)_0$, and reddening, $E({B-V})$, of the system to be $11.8{\pm}0.2$ mag and 0.45 mag, respectively. The mean proper motion of this system is $-5.37{\pm}0.81$ mas/yr and $0.31{\pm}0.40$ in RA and DEC directions, respectively. The mean proper motion of this system is found to be almost similar to the field region. The mass function for the brighter stars is found to be too high for the system to be an open cluster. These combined results place constraints on whether stellar system ESO65SC03 is a possible open star cluster remnant (POCR) or an Asterism. Our understanding is that the ESO65SC03 is in a stage of POCR by loosing their main sequence stars in the dynamic evolution processes.
\end{abstract}
\maketitle%
\section{INTRODUCTION}
\label{sec:intro}

The comprehensive study of the Open Star Clusters (OCs) and their basic parameters provide a fundamental database to understand the Galactic evolution processes. Due to the dynamic evolution processes such as stellar encounters, interactions with the Galactic tidal field, etc., a significant fraction of the cluster stars are gradually lost to the field regions. The main effects of dynamical interactions are mass segregation and evaporation of low-mass members \cite{and, pat, car1}. Studying the sparse region (the halo) along with the dense central region (the core) is therefore necessary to understand the evolution of any cluster. It is believed that the poorly populated OCs (PPOCs) do not survive longer than a few hundred Myr, whereas rich ones may survive for a longer period \cite{tan, car, bon}. The photometric and structural properties of such PPOCs are similar to those of Asterisms which sometime creates conflicting classifications \cite{bic2}. The PPOCs may have a RDP characterised essentially by a central overdensity together with significant noise outwards, typical features of an asterism. In these cases, only the CMD morphology may provide clues to the overdensity nature \cite{bic1}. In order to understand the difference between sparse clusters and asterisms as well as to find out any possible connection between them, we studied a poorly populated stellar system with a stellar enhanced region including PPOCs. For this, we searched PPOC systems in the WEBDA database and all available catalogues on VIZIER services. Kharchenko et al. \shortcite{kha} and Dias et al. \shortcite{dia} compiled several catalogues for optically visible and candidate OCs containing several basic spatial parameters of the cluster. We selected one such poorly populated stellar system ESO65SC03 for this study and extracted the data from various catalogues namely, 2MASS \cite{skr}, WISE \cite{wri}, SPM4.0 \cite{gir} and UCAC4 \cite{zac}.

The Two Micron All Sky Survey (2MASS; Skrutskie et al. 2006) has provided the photometric data of 410 million objects in J ($1.25~ \mu m$), H ($1.65 ~ \mu m$) and K ($2.17~\mu m$) bands. The limiting sensitivity of J, H, K bands with 3$\sigma$ are 17.1 mag, 16.4 mag and 15.3 mag, respectively. The $JHK-2MASS$ system gives more accurate isochrone fitting for the lower portions of CMDs which improve the turnoff points, distance modulii, ages and photometric membership of the clusters. Furthermore, this catalogue is a homogeneous database which allows us to observe young clusters in their dusty environments and reach out to their outer regions which is dominated by low mass stars \cite{tad}. The Wide-field Infrared Survey Explorer (WISE; Wright et al. 2010) is a NASA Medium Class Explorer mission that conducted a digital imaging survey of the entire sky in the mid-IR bands. The effective wavelength of mid-IR bands are 3.4 $\mu m$ ($W_1$), 4.6 $\mu m$ ($W_2$), 12 $\mu m$ ($W_3$) and 22 $\mu m$ ($W_4$). The SPM4.0 Catalogue provides position, absolute proper motions and photographic BV photometry for over a hundred million stars and galaxies  between the south celestial pole and -20 degrees declination and it is roughly complete down to V=17.5. It is also based on the photographic and CCD observations taken with the Yale Southern Observatory's double-astrograph at Cesco Observatory in El Leoncito, Argentina \cite{gir}. UCAC4 \cite{zac} is an all-sky star catalogue covering mainly 8 to 16 mag stars in a single passband between V and R. Furthermore, the positional errors are about 15 to 20 $mas$ for stars in the 10 to 14 mag range. The proper motion in this catalogue have been derived for around 113 million stars utilizing about 140 other star catalogs (including old SPM catalogue, Hipparcos and Tycho) with significant epoch difference to the UCAC4 CCD observations.\\

In the present paper, we derive the basic parameters and study the dynamic evolution of the stellar system ESO65SC03. This system is described as an optically visible sparce OSC in the Southern hemisphere \cite{kha}. The basic parameters and the dynamical properties of this system has never been studied in detail. Along with studying the dynamic evolution of this system, we also aim to find out whether it is a Possible Open Cluster Remnant or Asterism.
 
\section{SPATIAL STRUCTURE OF SYSTEM}
The classification of poorly populated stellar systems like ESO65SC03 is a challenging task due to the similarities in over-density between sparse OCs and asterisms. The estimation of structural parameters and dynamical properties of poorly populated stellar systems can be used as an excellent probe for their classification and to find out any evolutionary connection between them. Using the estimated radius in section 2.1, the extent of the system is drawn with a black circle in the W1 band of WISE image as illustrated in Fig.1. The detailed estimation of the system parameters are discussed in the following sub-sections.
\begin{figure}
\begin{center}
\includegraphics[width=21pc, angle=0]{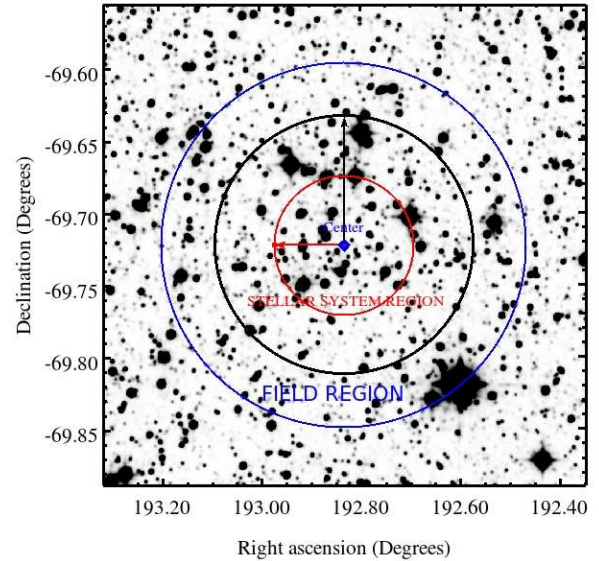}
\caption{The finding chart $(20'{\times}20')$ of stellar system ESO65SC03 in $W_1$ band of WISE taken from IRSA webportal. The field region starts after the radial distance of 5.36 arcmin (limiting radius shown in red circle) and ends in 7.58 arcmin (shown in blue circle). The region inside the black circle is the system region which is upto 5.36 arcmin radial distance (shown in black circle). }
\label{Fig1}
\end{center}
\end{figure}
\subsection{RADIUS OF STELLAR SYSTEM}
A structural study of the system ESO65SC03 has been carried out using those stars which were detected in each of the 2MASS bands (J, H and K) and their corresponding errors in magnitude $<$ 0.1 mag. These stars were used to identify the system coordinates using maximum stellar counts per unit area. The radial stellar densities were estimated by taking the concentric rings from the center of system with unit arcmin width, having maximum stellar counts and fitted with the King model \cite{kin}. If the fit was not perfect, we selected the next unit cell with maximum stellar counts until we achieved the best fit. Finally, we determined the center of the stellar system as $\alpha = 12^{h}:51^{m}:19.7^{s}, \delta = -69^{o}:43^{'}:21.6^{''}$ and the equivalent Galactic coordinate as $ l = 302.91^d, b = -6.84^d$ which is consistent with the Kharchenko et al. \shortcite{kha}.

The derived radial density profile (RDP) of the system is shown in Fig 2. The modified king empirical formula [$\rho_{r}- \rho_{0} = f_{0} (1+(r/r_{c})^2)^{-1}$, Kaluzny \& Udalski \shortcite{kal}] is drawn with the RDP of the cluster and it gives the peak stellar density ($f_{0}$), core-radius ($r_{c}$) and background stellar density ($\rho_{0}$) as $2.3{\pm}1.3~stars/arcmin^2$, $1.1{\pm} 0.6~ arcmin$, and $3.2{\pm}0.1~stars/arcmin^2$, respectively. Here, the core radius of any stellar system is the radial distance from its center where the enhanced stellar density above background {\bf level} becomes half of that of the center. The radius of the system is defined as the distance from the center of the system to the point where the stellar enhanced density is not separated from the field region. After adding the 1{$\sigma$} correction in the background stellar density, we found the radius to be $5.36{\pm}0.24$ arcmin.

Limiting radius of any stellar system is defined as a radial distance, after which no stars are gravitationally bound to the stellar system. It is independent of the background stellar density and only depends on peak stellar density, core radius and estimation in the error of the background stellar density. The Bukowiecki et al. \shortcite{buk} relation ($r_{lim} = r_{c} \sqrt{f_{0}/3 \sigma_{bg} -1}$) is used to estimate the limiting radius ($r_{limit}$) of the cluster that comes out to be $2.96{\pm}1.19~arcmin$, which is significantly smaller than the radius of the stellar system. Since, the system is gravitationally bound within $\sim 3$ arcmin and the model stellar radius is larger than expected gravitational attraction, therefore, the said stellar system may not be fall on the category of OCs. We found the {\bf spatial components} of this system as $X$ = 1.24 kpc, $Y$ = -1.92 kpc and $Z$ = -0.27 kpc. The Galactocentric distance of the cluster is computed to be about 7.51 kpc. Our estimated core-radius and system-radius are in close agreement with those determined by the Kharchenko et al. \shortcite{kha}.
\subsection{SURFACE NUMBER DENSITY PROFILE}
The number of stars per magnitude interval in unit area is defined as the surface number density (SND), $\sigma (r,m)$, and is a function of stellar magnitude and radius from the center \cite{sun}. We computed the $\sigma (r,m)$ for both the system and field regions as drawn in Fig.~3. We see that the $\sigma (r,m)$ of each region increases towards the higher magnitude end of each region. The SND profile of the system indicates that the system region have enhanced brighter stars compared to the fainter stars. The system stars are further divided into two groups through their SND profiles namely a) brighter stars group (BSG), b) fainter stars group (FSG). A separation at about 13 mag has been taken as cut-off magnitude between BSG and FSG group of stars. A vertical line separating the two groups for both the regions is shown in Fig.~3. The SND profiles of system and field regions give different slopes ($m=\frac{\Delta \sigma (r,m)}{\Delta M_{H}}$) for BSG and FSG which is estimated as $0.40{\pm}0.05$ and $-0.28{\pm}0.13$, respectively. For the field region, the slope of BSG and FSG are estimated as $0.60{\pm}0.11$ and $-0.07{\pm}0.09$, respectively. The slope of the BSG for the field region is found to be steeper than the system region while opposite is found for FSG.
\begin{figure}
\begin{center}
\includegraphics[width=13.5pc, angle=270]{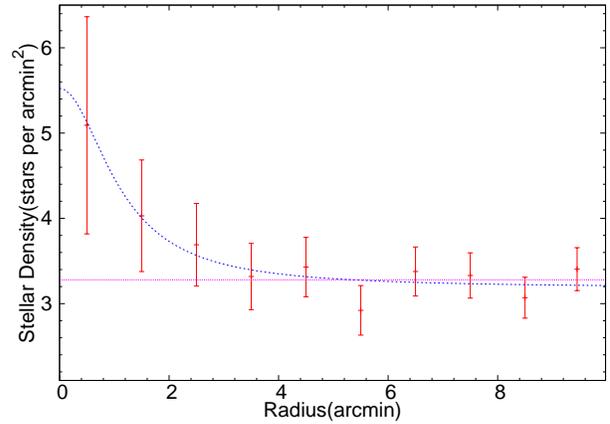}
\caption{The radial density profile (RDP) for the field of ES065SC03. The blue dashed line represents the best fit empirical profile on the cluster RDP. {\bf The pink line represents $\rho_{0}+\sigma_{\rho_{0}}$, where $\sigma_{\rho_{0}}$ is the error in the estimation of field stellar density ($\rho_{0}$).}}
\label{Fig2}
\end{center}
\end{figure}
\begin{figure}
\begin{center}
\includegraphics[width=13.5pc, angle=270]{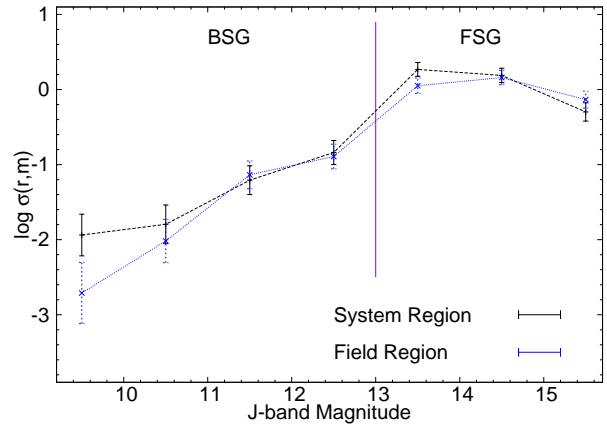}
\caption{The distribution of logarthim value of stellar number density versus $J$ magnitude. The black points represent the logarithm SND values for the system region while blue points represent same for the field region. The solid vertical line has been used to seperate the BSG and FSG stars of the stellar system.}
\label{Fig3}
\end{center}
\end{figure}

\begin{figure*}
\begin{center}
\includegraphics[width=20pc, height=40pc, angle=270]{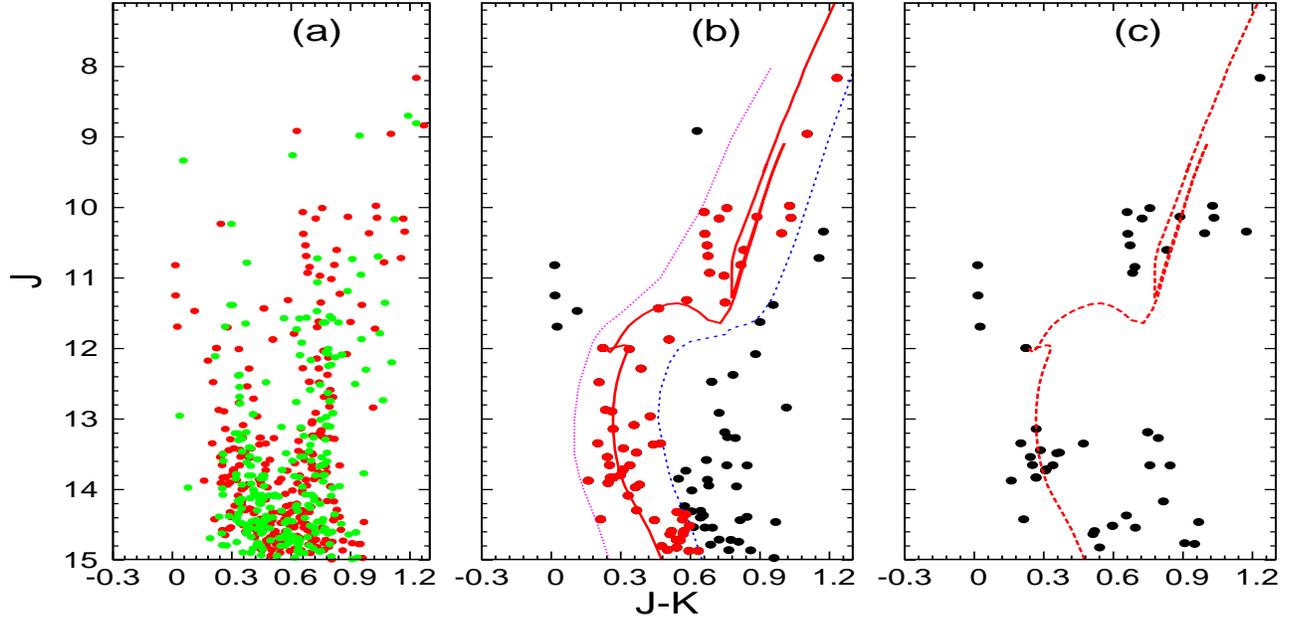}
\caption{J vs (J-K) CMD for the system. a) The cluster and the field stars are represented by red and green dots, respectively. b) CMD for the remaining stars after the field star subtraction through statistical method (grid size, $\Delta{J}=0.20$, $\Delta{(J-H)=0.05}$, $\Delta{(H-K)=0.05})$. The dashed lines represent the cluster evolutionary sequence while dotted lines represent the sequences beyond which CMDs may be contaminated by field stars. The black points represent possible residual field stars. c) CMD for the remaining stars after the field star subtraction through statistical method with increased grid size (${\Delta{J}=0.50, ~\Delta{(J-H)=0.20}, ~\Delta{(H-K)=0.20}}$) (see text for detail). }
\label{Fig4}
\end{center}
\end{figure*}

\subsection{COLOUR-MAGNITUDE DIAGRAM (CMD)}
There is a possibility of some stars getting rejected from the system membership, if the choosen field region has more area than the system region. Whereas, if the area of the choosen field is less, the probablity of the field stars in the stellar system is increased. In either case, the observed CMD must get deformed, thereby increasing the inaccuracy in estimation of the system parameters. Therefore, we have considered the area of the field region as 90.29 $arcmin^2$ which is equal to the area of the stellar system. Our field region begins after the radial distance of 5.36 arcmin from the centre as shown in Fig 1. The field star decontamination is necessary for reducing the color-scattering on the CMDs. The field stars are removed by comparing the magnitude and colour similarities of stars between the stellar system to the field region. Based on the following criteria, the stars are excluded from the stellar enhanced region:

\begin{itemize}
\item The obtained J-band magnitude difference between star in the system is not more than of 0.20 mag from the star of the field region at the same place in the CMD.
\item The colour difference ($J-H$) of stars between system and field regions is not more than 0.05 mag.
\item Similarly, the colour difference ($H-K$) of stars between system and field regions is not more than 0.05 mag.
\end{itemize}

We define a term $D_{mc}= \sqrt {\Delta_{J}^2+ \Delta_{JH}^2 + \Delta_{HK}^2}$ which is the summation of color and magnitude differences of the field star and the star in the system, which fall within the grid of above specifications. We removed the star from the grid whose $D_{mc}$ is minimum. After iterating the procedure for each field region star, we found 112 stars on the statistically cleaned CMDs.

The statistical approach depends on the color and magnitude grid size. If the boundaries of the grid size is increased, the system stars may get excluded from the membership of the stellar system and if the grid size is decreased, there is a possibility for the field stars to remain in the stellar system. Thus, the optimum grid size is chosen in such a way that the main sequence pattern appears in the observed CMD. Before the field star decontamination, no such pattern was visible while after the decontamination of the field stars, we see a clear main sequence pattern (Fig 4-b). {\bf We noticed an almost vertical sequence of stars with $(J-K){\sim}0.8$ and $J$ fainter than 13 mag in the $2MASS$ decontamination CMD. This vertical sequence represents residual field star contamination. These residual field stars are removed from the decontaminated CMD by taking grid size as $\Delta{J}=0.5,~\Delta{(J-H)}=0.2~{\&}~ \Delta{(H-K)}=0.2$ but exact cluster evolutionary sequence is not occurred in the decontaminated CMD (Fig 4-d). In this scenario, the estimation of system parameters may not be possible, therefore, we have derived parameters through decontaminated CMD as obtained by applying the earlier grid size.} The limitation of this approach, however, is that the cluster characteristics depend on the size of the grid.

The observed CMDs of the stellar system are used to determine the model distance, age and reddening of the system by comparing them with the theoretical stellar isochrones. To test a suitable number of possible isochrone solutions, we first determined initial values for each physical parameter by visual inspection [$log(\tau)_i,~(m-M)_{i},~(J-K)_i$ ] and subsequently we investigated a large parameter space in an iterative manner. Our isochrone solutions typically span the physical ranges as $log(\tau)=log(\tau)_i~{\pm}~0.05,~(m-M)=(m-M)_{i}~{\pm}~0.50$ and $(J-K)=(J-K)_i~{\pm}~0.10$. We used the solar metallicity isochrones of log(age)=8.75 of $UBVRIJHK$ photometric system \cite{mar} in $J$ vs $(J-K)$ CMD (Fig 4-b). Through these CMDs, we found the apparent distance modulus of J-band i.e. $(m-M)_{J}$, and log(age) of the system as 12.06 mag, and 7.75 Myr, respectively. Furthermore, the color-excess, $E({J-K})$, is estimated as $0.22$ mag. The obtained log(age) value is in close agreement with that given by Kharchenko et al. \shortcite{kha}. The obtained age of stellar system is however higher than that of sparse OCs \cite{bon, tan}.

The Fiorucci \& Munari \shortcite{fio} relations ($\frac{E({J-H})}{E({B-V})}= 0.309 {\pm} 0.130, ~\frac{E({J-K})}{E({B-V})} = 0.485 {\pm} 0.150$) have been used to estimate the reddening i.e. $E({B-V})$ and $E({J-H})$, which are estimated as 0.45 mag and 0.14 mag, respectively. The corresponding absolute distance modulus [$(m-M)_{0}= (m-M)_{J}-1.125E({J-K})$] and distance obtained as $11.81{\pm}0.25$ mag and $2.30{\pm}0.35$ kpc, respectively. The error in these values are estimated by considering the errors of Fiorucci \& Munari \shortcite{fio} relations. Kharchenko et al. \shortcite{kha} estimated the apparent distance modulus, reddening and distance as 12.6 mag, 0.56 mag and 3.05 kpc, respectively which is slightly different than our obtained values.

\subsection{TWO COLOUR DIAGRAMS (TCDs)}
Two color diagrams are useful to derive the relationships among various colours. The color ratio, $\frac{J-K}{V_{ph}-K}$ and $\frac{J-K}{(B-V)_{ph}}$, are obtained as $0.19{\pm}0.01$ mag and $0.30{\pm}0.03$ mag, respectively and the resultant plots are shown in Fig 5(a \& b). The $(V_{ph}-J)$ vs $(J-K)$ TCD are utilized to estimate the interstellar extinction in the near-IR. The best fit in this TCD gives a colour-excess of $E(V_{ph}-K)=0.90{\pm}0.05$ mag and $E(J-K)=0.17{\pm}0.19$ mag. We used Whittet and van Breda \shortcite{whi} relation 
to estimate the reddening vector, $R_{cluster}$=1.1$E(V-K)/E(B-V)$ in the direction of system. We have found this vector as 2.2 which is far-off from the normal value of 3.1. Dutra et al. \shortcite{dut} relation $\frac{A_K}{A_V}=0.112$ yields the $A_{K}$ as 0.11. In fact, normal reddening law ($R_{V}$=3.1) is expected in the line of sight of the cluster in the absence of dust and intermediate stellar gases \cite{sne}. So, we have concluded that system region is affected by interstellar gases and dust.
\begin{figure}
\begin{center}
\includegraphics[width=13.0pc, angle=270]{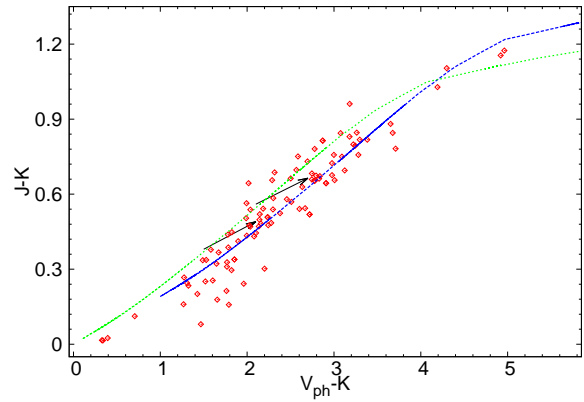}
\caption{The ($V_{ph}-K$) vs ($J-K$) TCD for statistically cleaned members of system. The green dotted line shows the solar metallicity isochrone for log(age)=8.75, while the two solid arrows indicate the direction of the normal reddening vector. The blue dotted line is obtained by using reddening as $E(V-K)=0.90$ mag and $E(J-K)=0.17$ mag.}
\label{Fig5}
\end{center}
\end{figure}
\begin{figure}
\begin{center}
\includegraphics[width=21pc, height=30pc, angle=0]{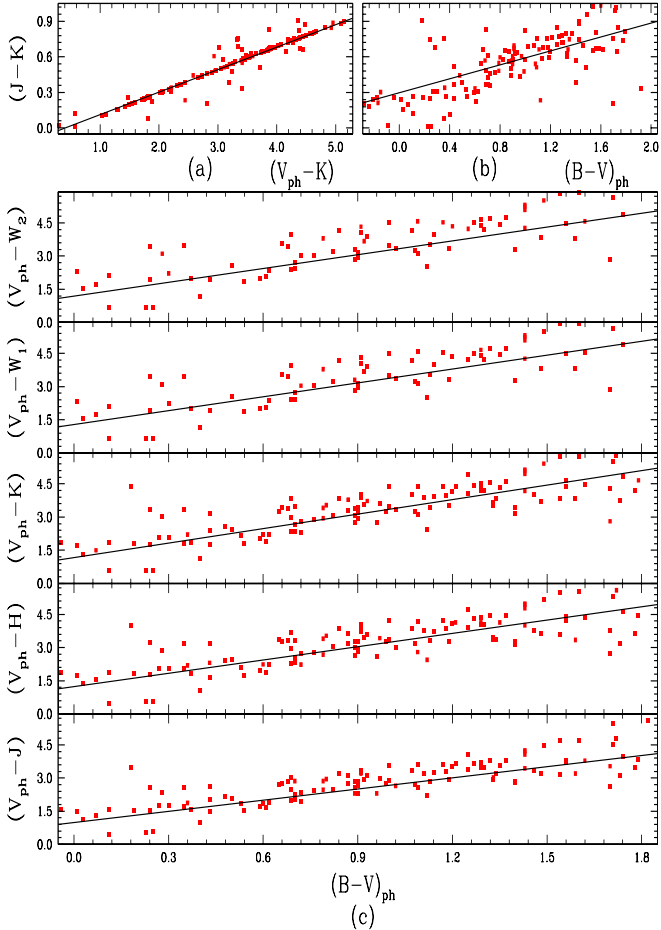}
\caption{The ($V_{ph}-K$) vs $(J-K)$ and $(B-V)_{ph}$ vs $(J-K)$ (a \& b) TCDs for the system's members which are found SPM catalogue. The dotted straight line represents the best linear fit. The lower panel represents the $(V_{ph}-\lambda)/(B-V)$ diagrams where $\lambda$ have indicated any one photometric band from $J$, $H$, $K$, $W_1$ and $W_2$ bands.}
\label{Fig6}
\end{center}
\end{figure}
\begin{table}
\caption{The various color ratios are given in the table. The third column represents the estimated normal colour ration through $R_{cluster}$ and Neckel and Chini relation \shortcite{nec} while the value in fourth column represents the expected normal colour ratio values.}
\begin{center}
\begin{tabular}{@{}ccccc@{}cc}
\hline\hline
Colour && Obtained && Estimated && Expected\\
Ratio  && slope && normal slope && normal value\\
\hline%
 && && \\
 $\frac{V_{ph}-J}{(B-V)_{ph}}$  && $1.34{\pm}0.11$ && $1.89{\pm}0.19$ && 1.96\\
 && && \\
 $\frac{V_{ph}-H}{(B-V)_{ph}}$  && $1.52{\pm}0.13$ && $2.14{\pm}0.27$ && 2.42\\
 && && \\
 $\frac{V_{ph}-K}{(B-V)_{ph}}$  && $1.66{\pm}0.14$ && $2.34{\pm}0.32$ && 2.60\\ 
 && && \\
 $\frac{V_{ph}-W_1}{(B-V)_{ph}}$  && $1.69{\pm}0.20$ && $2.38{\pm}0.48$ && - \\ 
 && && \\
 $\frac{V_{ph}-W_2}{(B-V)_{ph}}$  && $1.68{\pm}0.20$ && $2.37{\pm}0.47$ && - \\ 
\hline\hline
\end{tabular}
\end{center}
\label{tab1}
\end{table}

Further, the TCD are used to examine the nature of variation of color-excess with increasing wavelength and normal colour ratio values in the direction of system. For this purpose, the $(V_{ph}-\lambda) $ vs $(B-V)_{ph}$ ($\lambda$ is one of $J,~ H, ~K, ~W_1$ and $W_2$ bands.) plots are constructed in Fig 5(c) and the resultant colour ratios ($m_{cluster}$) are listed in Table 1. This clearly indicates that the value of $m_{cluster}$ increases from $J$ to $W_1$. To derive the normal values for various colour-ratios, we have used the following approximate Neckel {\&} Chini \shortcite{nec} relation,
$$m_{normal}=m_{cluster}{\times}\frac{R_{normal}}{R_{cluster}}$$
The present estimated normal values are close to known values. The variation of these values may be occurred because photographic plate are less sensitive to measurement of stellar magnitudes and have more errors.
 
\subsection{PROPER MOTION OF THE SYSTEM}
We found a total of 158 stars in statistical cleaned J vs (J-K) CMD. We have also drawn a {\bf colour-magnitude filter} around the best fitted isochrone as shown in Fig.~4(b) by dotted lines. A total of 73 stars are found inside the {\bf colour-magnitude filter}, which are defined as the main sequence (MS) stars and considered for the proper motion study (depicted by red dots in Fig 4-b). The mean proper motion of this stellar system is estimated using two different catalogues. Using a criterion of 1 arcsec uncertainity, a total of 62 and 69 MS stars matched with the UCAC4 and SPM4.0 catalogues, respectively. The stars which do not fall within $3~\sigma$ value of the mean are rejected. Using proper motions values from UCAC4 catalogue for 59 remaining MS stars after $3~\sigma$ clipping, the mean proper motion values of the stellar system is found to be ${\bar{\mu}}_{x}$ = $-2.42{\pm}0.99$ mas/yr, and  ${\bar{\mu}}_{y}$ = $-2.18{\pm}0.78$ mas/yr in RA and DEC directions, respectively. Similarly, the mean proper motion values of remaining 61 MS stars, as found through SPM4.0 catalogue, are found to be $ -5.37 {\pm} 0.81$ mas/yr and $ 0.31{\pm}0.40$ mas/yr. One can see that both the catalogues have different number of stars and give different values of mean proper motion. Since, the uncertainties in the stellar proper motion values of the SPM4.0 catalogue is less than UCAC4 catalogue, therefore, we considered the mean proper motion of the system ${\bar{\mu}}_{x}$ = $-2.42{\pm}0.99$ mas/yr, and  ${\bar{\mu}}_{y}$ = $-2.18{\pm}0.78$ mas/yr in RA and DEC directions, respectively as obtained through the former catalogue. Our obtained proper motion do not match with that estimated by Kharchenko et al. \shortcite{kha} using the PPMXL catalogue (${\bar{\mu}}_{x}$=-14.54 mas/yr, ${\bar{\mu}}_{y}$=4.63 mas/yr).
\begin{figure}
\begin{center}
\includegraphics[width=16.0pc, angle=270]{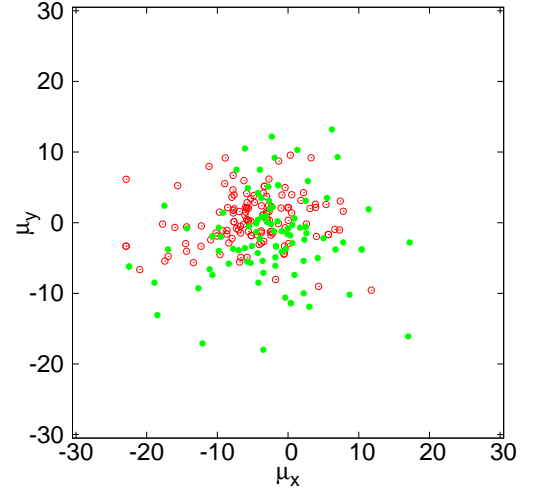}
\caption{The open circles are depicted the location of stars at $\mu_{x}-\mu_{y}$ plane based on the proper motion of the system in SPM4.0 catalogue for the member of the stellar system. The filled circles represent the same using the UCAC4 catalogue.}
\label{Fig7}
\end{center}
\end{figure}

When we consider all the stars in Fig4(b) without any colour-magnitude filter, and adopting the same approach as above using the SPM4.0 catalogue, we found a mean proper motion of  ${\bar{\mu}}_{x} = -4.96 {\pm} 0.56$ and ${\bar{\mu}}_{y} = 0.43{\pm}0.32$ for the system. This value is similar to the mean proper motion ${\bar{\mu}}_{x} = -4.45{\pm}0.32$ mas/yr and ${\bar{\mu}}_{y} = 0.48{\pm}0.23$ mas/yr estimated for the field region. For gravitationally bound cluster system, this result is unexpected. Hence, we could not separate out the stars between the system and field regions on the basis of proper motion membership probabilities.

\section{DYNAMICAL STUDIES OF SYSTEM}

The dynamical properties provide the information about the member evaporation rate, moving direction of cluster members and the strength of gravitational interaction within systems. In order to understand the dynamical behavior of OCs, the mass function and mass segregation within the cluster are looked into. We studied the dynamical properties of the system by assuming two sample of stars. The first SAMPLE-A contains those stars which are found on statistically cleaned near-IR CMDs. The second SAMPLE-B consists of those members which are common among SAMPLE-A, 2MASS catalogue (near-IR survey) and WISE catalogue (mid-IR survey) within 1 arcsec uncertainty. Generally the brighter stars are found in almost all filters but detection of fainter stars varies from filter to filter. This is because the completeness factor of stars gradually decreases towards the fainter end of stellar magnitude, as indicated by several completeness studies for different filters in the past \cite{jos}. By studying two different samples, we can understand the effect of stellar incompleteness on dynamical behavior of the system. Table 2 lists all the stars of SAMPLE-A and SAMPLE-B in various H-band magnitude bins. The analysis of dynamical study of these samples is described below.

\subsection{MASS FUNCTION}

During the process of star formation, the distribution of stellar mass per unit volume is defined as the initial mass function (IMF) which determines the subsequent evolution of the cluster \cite{kro}. The direct estimation of the IMF is not possible due to the dynamical evolution of the cluster. However, we can derive the mass function (MF), which is the relative number of stars per unit mass and can be expressed by a power law $N(\log M) \propto M^{\Gamma}$. The slope, $\Gamma$, of the MF can be determined from
$$\Gamma = \frac{d\log N(\log\it{m})}{d\log\it{m}}$$
\noindent where $N\log(m)$ is the number of stars per unit logarithmic mass. The masses of probable cluster members can be determined by comparing observed magnitudes with those predicted by a stellar evolutionary model
if the age, reddening, distance and metallicity are known. The MF values estimated for the cluster are listed in Table 2, while the variation of MF with average mass in different magnitude bins are shown in Fig. 8. 

\begin{table}
\caption{The MF of the cluster ESO 65 03 derived from cleaned (H-K)/H CMD through (a) stars from 2MASS catalog and (b) common stars between WISE and 2MASS (represented by suffix A and B respectively).}
\begin{center}
\begin{tabular}{@{}cccccc@{}}
\hline\hline
H range & Mass range & $\bar{m}$ & $log(\bar{m})$ & $N_{A}$ & $N_{B}$\\
(mag) & $M_{\odot}$  & $M_{\odot}$ & & &\\
\hline%
 08-09  & 2.567-2.564 & 2.566 & 0.409 & 02 & 02\\
 09-10  & 2.564-2.561 & 2.562 & 0.409 & 11 & 10\\
 10-11  & 2.561-2.553 & 2.557 & 0.408 & 10 & 09\\
 11-12  & 2.553-2.438 & 2.495 & 0.397 & 15 & 11\\
 12-13  & 2.438-2.058 & 2.248 & 0.352 & 14 & 14\\
 13-14  & 2.058-1.500 & 1.779 & 0.250 & 43 & 29\\ 
 14-15  & 1.500-1.112 & 1.306 & 0.116 & 30 & 14\\
 15-16  & 1.112-0.850 & 0.981 & -0.008 & 29 & 05\\ 
\hline
\end{tabular}
\begin{tabular}{@{}ccccc@{}}
\hline
H range & $log(\Phi)_{A}$ & $e_{log(\Phi)_{A}}$ & $log(\Phi)_{B}$ & $e_{log(\Phi)_{B}}$\\
\hline%
 08-09  & 3.595 & 0.707 & 3.595 & 0.707 \\
 09-10  & 4.335 & 0.302 & 4.294 & 0.316 \\
 10-11  & 3.867 & 0.316 & 3.821 & 0.333 \\
 11-12  & 2.875 & 0.258 & 2.740 & 0.302 \\
 12-13  & 2.279 & 0.267 & 2.279 & 0.267 \\
 13-14  & 2.496 & 0.152 & 2.325 & 0.186 \\ 
 14-15  & 2.363 & 0.183 & 2.032 & 0.267 \\
 15-16  & 2.395 & 0.186 & 1.632 & 0.447 \\ 
\hline\hline
\end{tabular}
\end{center}
\label{tab1}
\end{table}

\begin{figure}
\begin{center}
\includegraphics[width=13.0pc, angle=270]{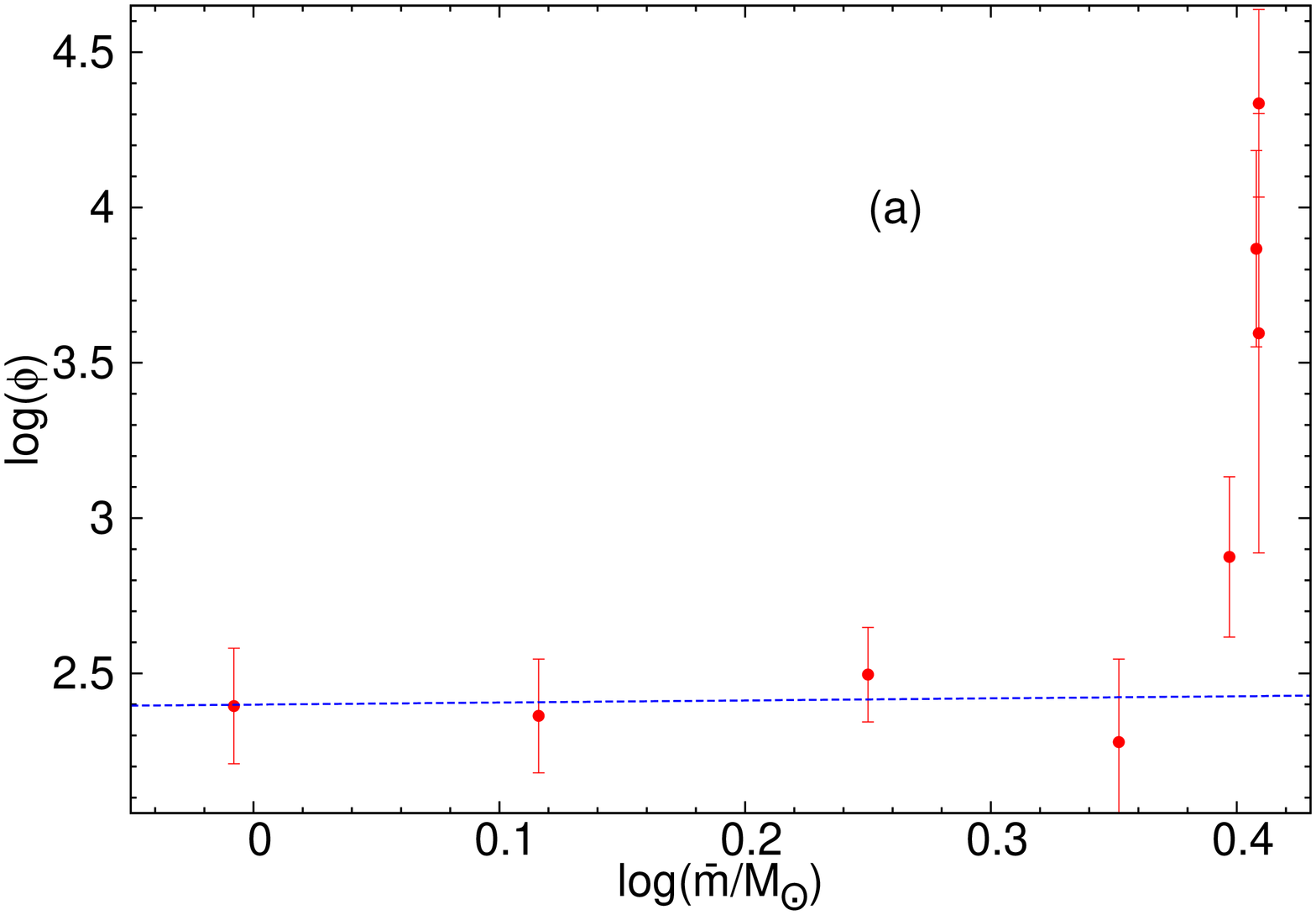}
\includegraphics[width=13.0pc, angle=270]{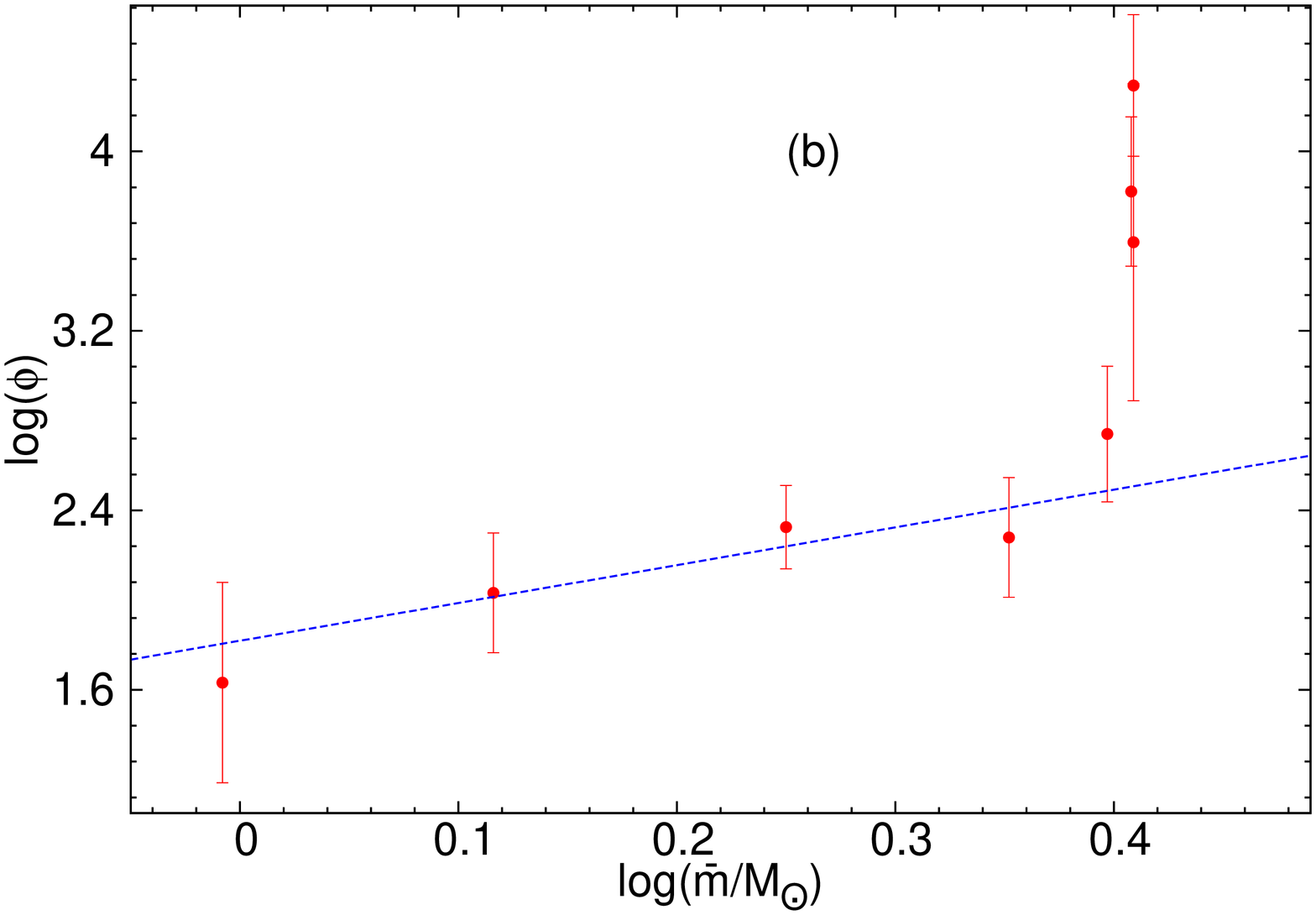}
\caption{The variation of logarithm MF with the logarirhm of average mass of members. These quantities have been comuted using members as leftover after field star {\bf decontamination} through the statistical procedure. The figure A and B are represented the said plot for members of SAMPLE-A and SAMPLE-B.}
\label{Fig8}
\end{center}
\end{figure}
 
To determine the MF slope, we have only used main sequence stars which having H-magnitude greater than 12 mag. The mostly brighter stars may be giants members of systems, therefore, they have excluded to estimation of MF slope. The MF slope value $\Gamma$ is found to be $0.07{\pm}0.42$ for the mass range from 2.44 $M_{\odot}$ to 0.85 $M_{\odot}$ which is low numerical value compared to the Salpeter value [Salpeter \shortcite{sal} gave the classical slope of -1.35 for the stars of mass range 0.4 $M_{\odot}$ to 10 $M_{\odot}$] and have positive slope. Such flatness/positive value of MF slope may be due to the dynamical evolution, rather than the intrinsic difference in the initial MF \cite{nil}. The stellar incompleteness increases toward the fainter end of photometric magnitude and consequently reduce the MF value thereby increasing the flatness of the MF slope. The MF values have also been computed for SAMPLE-B members of stellar system. The purpose of computing the MF values for SAMPLE-B members was to find out any correlation between MF values and the common stellar fraction of stars in various photometric regions. The MF value of the stars from SAMPLE-B increases as the average mass of the stars increases from 0.85 $M_{\odot}$ to 2.57 $M_{\odot}$. Brighter stars are vital members to identify any cluster. However, even after the cleaning, the field stars may still be present in the fainter end. The MF value depends on the number of stellar members, hence the MF value should increase in presence of the residual field stars. However, the stellar incompleteness gives the opposite result, which is much effective and prominent compared to the effect of residual field stars.
The corresponding MF slope value for SAMPLE-B is $1.68{\pm}0.67$ which is numerically comparable with the Salpeter value but opposite in slope. Since, the stars of SAMPLE-B are found in both 2MASS and WISE catalogues, the common stellar fraction of stars in both catalogues gradually decreases towards the end of fainter stars. Due to the incompleteness of the data, in terms of the number of common stars, we expect the MF slope increases towards the positive slope value.

\subsection{MASS SEGREGATION}

Through the process of Mass segregation, the stellar encounters of the cluster members are gradually increased. Furthermore, the higher-mass cluster members gradually sink towards the cluster center by transferring their kinetic energy to the more numerous lower-mass stellar members of the cluster \cite{yad}. For the investigation of the dynamical evolution and mass segregation phenomenon due to the energy equipartition, the cluster stars are divided into two mass range of $2.567 ~{\leq} M_{\odot} < 2.438$ and $2.438 ~{\leq} M_{\odot} < 0.850$. The cumulative radial stellar distribution of stars for various mass ranges are shown in Fig 8. This clearly indicates that the mass segregation effect is present in the stellar system, and the radial commutative distributions of various mass groups are seen separately from system center to a distance of about 4.2 arcmin from the centre. The Kolmogorov-Smirnov (K-S) test have been performed for verifying the same kind of mass distribution. Using the relation $\chi^2=\frac{4 D^2 n_1 n_2}{n_1 + n_2}$ we have also estimated the $\chi^2$ value of mass segregation for the system, where D is the maximum distance between two cumulative radial distribution of stars and $n_1$ and $n_2$ are the number of stars found for various mass range. For the SAMPLE-A \& SAMPLE-B the $\chi^2$ value comes out to be 4.14 (confidence level is 87\%) and 3.38 (confidence level is 81\%), respectively. On the basis of this K-S test, we conclude that mass segregation phenomenon is more for the stars of SAMPLE-A than the other sample.

\begin{figure}
\begin{center}
\includegraphics[width=13pc, angle=270]{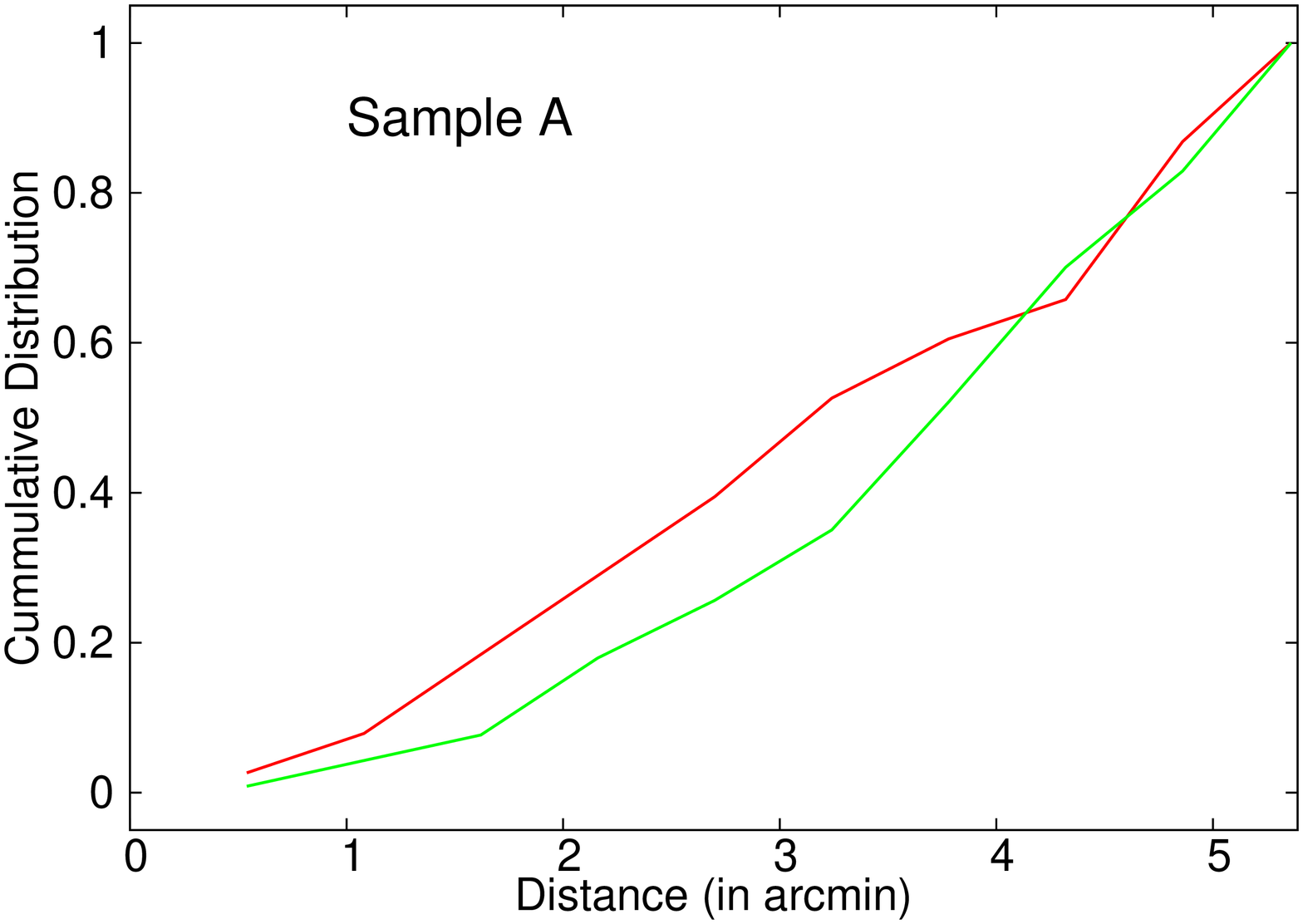}
\includegraphics[width=13pc, angle=270]{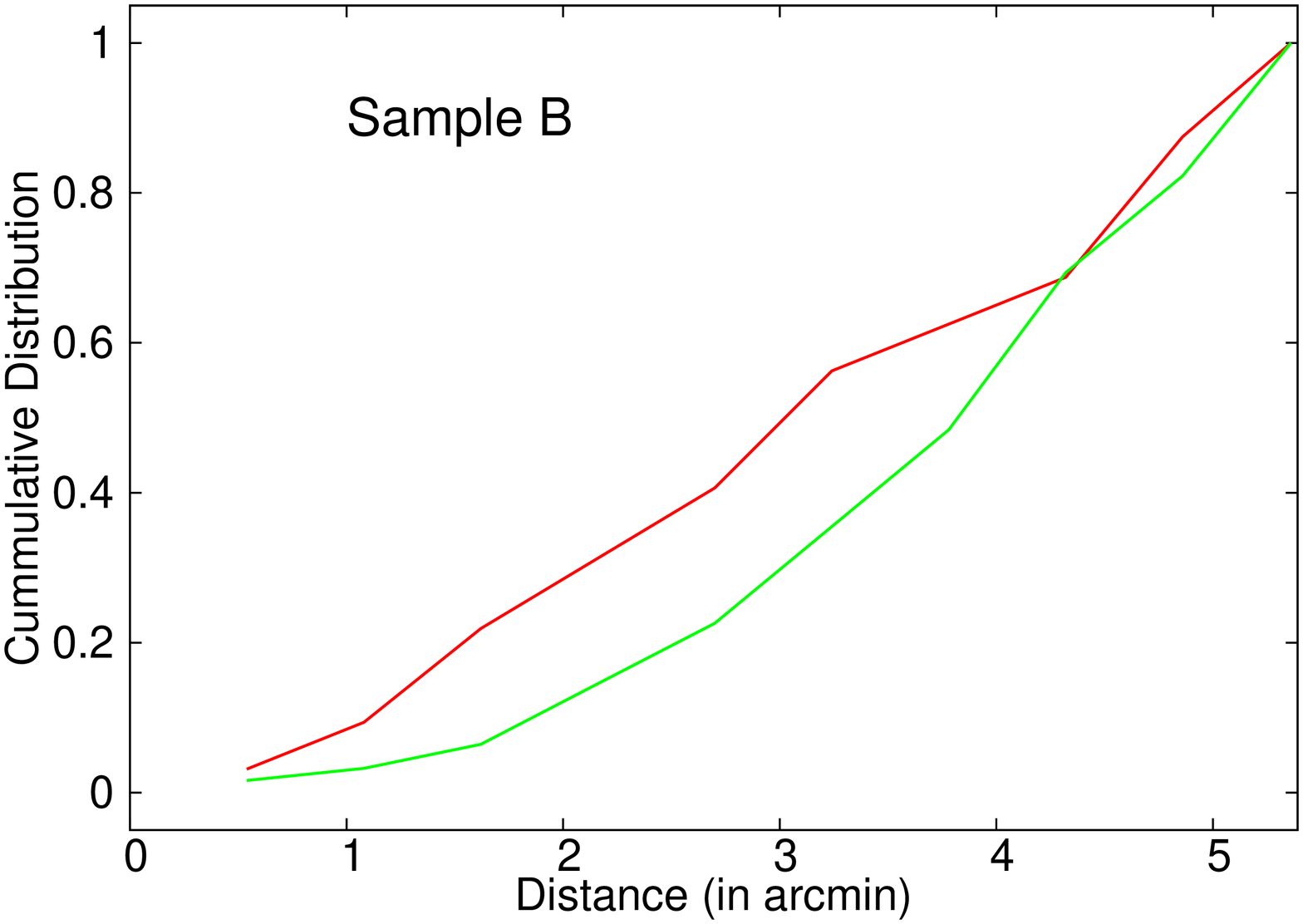}
\caption{The radial density cumulative distribition of stars for various mass groups. The mass segregation phenomenon for the star groups are shown. The red and green lines represent the mass range : $2.567 ~{\leq} M_{\odot} < 2.438$ and $2.438 ~{\leq} M_{\odot} < 0.850$ respectively.}
\label{Fig9}
\end{center}
\end{figure}

\begin{table*}
\caption{The structural and dynamical parameters of stellar system have been listed here. In this table, G.D. and H.D. represent Galactocentric distance and Heliocentric distance respectively, while other notations are described in text.} 
\begin{center}
\begin{tabular*}{\textwidth}{@{} c\x c\x c\x c\x c\x c\x c\x c\x c@{}}
\hline \hline
  $G.D.$   & $H.D.$     &  $X$  &  $Y$ 
         & $Z$ & $(m-M)_{J}$     & $E({J-K})$ 
          & $E({B-V})$  & MF slope \\
 (kpc)        & (kpc)          & (kpc)                & (kpc)
         & (kpc)                          & (mag)
         & (mag)           & (mag)       & - \\
\hline
 7.51 & $2.30{\pm}0.35$ & 1.24 & -1.92 & -0.27 & 12.06 & 0.22  & 0.45 & $0.07{\pm}0.42$ \\ 
\hline \hline
\end{tabular*}\label{tab2}
\end{center}
\end{table*}

\section{DISCUSSION}
The PPOCs and Asterisms have similar structural properties \cite{bic2} but only differ in their comprehensive structural and dynamical studies. The over-density profile does not provide any clue for their type. However, SNDP may provide us with an in-depth knowledge of their charactericstics. The Figs.~4(b) and 4(c) show the dependency of observed MS on Grid size of statistically cleaned procedure. The smaller grid size of $\Delta_{J}=0.20, ~\Delta_{JH}=0.05, ~\Delta_{HK}=0.05$ provides the MS pattern for the stellar system as well as the isochrone solution. We do not find any MS pattern on the observed CMD when the grid size is increased to $\Delta_{J}=0.50, ~\Delta_{JH}=0.20, ~\Delta_{HK}=0.20$ which is less than the typical grid sizes of sparse clusters. Thus the grid size did not provide any useful information about the MS characteristics of the stellar system.

Using two different catalogues [PPMXL \cite{ros} and UCAC4 \cite{zac}], Joshi et al. \shortcite{jos} found that the proper motion of the OC NGC 559 is similar to each other within their given uncertainities. Due to the low subtraction efficiency of the field decontamination procedure, the mean proper-motion of the system may get shifted towards the proper motion centre of the field regions, as expected. The mean proper motion value of the field and the system regions, as estimated by us, are almost similar. As a result, the low subtraction efficiency of the field decontamination have negligible effect on the mean proper motion of the system. The stellar density of the system will increase due to the low subtracting efficiency, leading to the increased MF but the slope value of MF decreases due to the stellar incompleteness. In our study, the effect of later is more dominant than the low subtracting efficiency. However, if the corrected completeness factor is applied to the stellar system, the slope of the MF may increase. Due to fainter members of ESO65SC03 we also found the flat MF slope indicating more evolved members. In addition, the MF values of brighter members of the system is too low for it to be an OC. The typical MF value for brighter members of any OC reported is much lower than our value \cite{lat, jos, glu, yad, tad1, tad, bha, sha}. There is also a possiblity that the lighter members of this system has already evaporated from its central region as expected by the phenomenon of Mass Segregation.

\section{CONCLUSION}
The  stellar enhancement over the field region is clearly visible in the RDP of ESO65SC03 and the enhancemnt in the SNDP is only due to the brighter members of the system. Since the SNDPs of the field and system regions are intersecting each other at several points, the system does not seem to be an open star cluster. On the other hand, the limiting radius is found to be significantly smaller than the cluster radius which is unexpected for an OC system. We estimated the age, heliocentric distance, galactocentric distance and reddening of the system as 562 Myr, $2.30{\pm}0.35$ kpc, 7.51 kpc and 0.45 mag, respectively.

The Two-Color diagrams for this system indicate that the color excess values [$E({V_{ph} -\lambda})$] increases from J-band to $W_1$-band. Our photographic color-ratio values for the stellar system is less than the typical [$E({V_{CCD}-\lambda})$] values found through CCD observations. The mass function value for the brighter stars is too low for the system to be an OSC. However for stars with mass $2.57<M_{\odot}<0.85$ the MF slope is found to be $0.07{\pm}0.42$ indicating that most of the stars are evolved from the MS. The MF slope for the common stars between WISE and 2-MASS catalogue is $1.68{\pm}0.67$ which indicates that the MF slope increases towards the positive value due to the incompleteness effect. The system also shows the mass segregation phenomenon. The number of stars in the system are found to be about 6.5\% more than that of the field region. Based on all the above said factors and considering the criteria suggested by Pavani et al. \shortcite{pav}, we conclude that ESO65SC03 may not be a normal star cluster but a possible open cluster remnants (POCR). Since, we get an isochrone solution for system by applying limiting grid size in statistical field star decontamination procedure, therefore, the possibility of system as a Asterism is ruled out. We conclude that ESO65SC03 is reached at a stage of POCR by losing their MS stars in the dynamic evolution processes through the tidal effects and stellar encounters. A further multi-bands analysis is required to conclude the exact nature of this system.

\begin{acknowledgements}
GCJ is thankful to University Grants Commission (New Delhi) for the financial assistance in terms of a RFSMS fellowship. SC acknowledges the Indo-Russian (RFBR) project INT/RFBR/P-118 for the financial support. Authors are also thankful to the referee for his constructive comments and suggestions which helped us to improve the scientific contents. This publication made use of data products from the Two Micron All Sky Survey (2MASS), a joint project of the University of Massachusetts and the Infrared Processing and Analysis Center/California Institute of Technology, funded by the National Aeronautics and Space Administration and the National Science Foundation. This publication used the data products from the Wide-field Infrared Survey Explorer (WISE), which is a joint project of the University of California, Los Angeles, and the Jet Propulsion Laboratory/California Institute of Technology, funded by the National Aeronautics and Space Administration.
\end{acknowledgements}


\end{document}